\documentclass[conference]{IEEEtran}
\IEEEoverridecommandlockouts
\usepackage{cite}
\usepackage{amsmath,amssymb,amsfonts,amsthm}
\usepackage{algorithmic}
\usepackage{graphicx}
\usepackage{textcomp}
\usepackage{xcolor}
\usepackage{typearea}
\def\BibTeX{{\rm B\kern-.05em{\sc i\kern-.025em b}\kern-.08em
    T\kern-.1667em\lower.7ex\hbox{E}\kern-.125emX}}

\newcommand{\stsets}[1]{\mathbb{#1}}
\newcommand{\R}{\stsets{R}}
\newcommand{\M}{\stsets{M}}

\newcommand{\N}{\stsets{N}}

\theoremstyle{definition}

\theoremstyle{remark}

\newtheoremstyle{mytheorem}{0.5cm}{0.2cm}{\slshape}{ }{\bfseries}{.}{ }{}
\theoremstyle{mytheorem}

\renewcommand{\P}{\mathbf{P}}

\DeclareMathOperator{\E}{{\bf E}}

\DeclareMathOperator{\one}{{ 1\hspace*{-0.55ex}I}}
\newcommand{\cond}{\hspace*{1ex} \rule[-1ex]{0.15ex}{3ex}\hspace*{1ex}}

\newcommand{\thru}{,\dotsc,}

\DeclareMathOperator{\argmin}{arg\,min}

\renewcommand{\epsilon}{\varepsilon}

\renewcommand{\phi}{\varphi}

\newcommand{\ssp}{\hspace{0.5pt}}
\newcommand{\seg}{see, \hbox{e.\ssp g.,}}
\newcommand{\ie}{\hbox{i.\ssp e.}}

\newcommand{\iid}{\hbox{i.\ssp i.\ssp d.}}

\newcommand{\deq}{\overset{\mathcal{D}}{=}}

\newcommand{\Sib}{\ensuremath{\mathsf{Sib}}}

\newcommand{\sas}{\ensuremath{\mathrm{St\alpha S}}}
\newcommand{\tas}{\ensuremath{\mathrm{T\alpha S}}}
\newcommand{\das}{\ensuremath{\mathrm{D\alpha S}}}

\newlength{\querylen}
\usepackage{fancybox}

\begin{document}

\title{Thinning-Stable Point Processes as a Model for Spatial
  Burstiness\\ \medskip
\textit{\small{WiOpt-2025 Conference paper, 26-29th of May, 2025, Linköping, Sweden }}}

\author{\IEEEauthorblockN{%1\textsuperscript{st}
    Sergei Zuyev\IEEEauthorblockA{\textit{Chalmers University of Technology and
        University of Gothenburg,}\\
        \textit{Department of Mathematical Sciences},\\ 412
    96 Gothenburg, Sweden.\\ Email: sergei.zuyev@chalmers.se}}}

\maketitle

\begin{abstract}
  In modern telecommunications, spatial burstiness of data traffic
  poses challenges to traditional Poisson-based models. This paper
  describes application of thinning-stable point processes,
  which provide a more appropriate framework for modeling bursty
  spatial data. We discuss their properties, representation, inference
  methods, and applications, demonstrating the advantages over
  classical approaches.
\end{abstract}


\section{Introduction}
The classical Poisson point process has been a foundational model in
queueing theory for telecommunications since the early 20th century
\cite{erlang1909, palm1943}. It assumes that events (such as call
arrivals) occur independently, with a constant or slowly varying mean
rate, \seg~\cite{Kl_roc:75}. However, real-world data, particularly in
mobile communications, often exhibit burstiness—periods of high
activity interspersed with inactivity. The activity rate can change
rapidly by orders of magnitude; for instance, the size of a typical
SMS message or email is several orders of magnitude smaller than that
of a streamed video.

Traditional Poisson models struggle to capture this burstiness,
leading to inefficient network resource allocation and inaccurate
performance predictions. While heavy-tailed queueing models and
fractional Brownian motion have been employed to account for temporal
burstiness, they are poorly suited for modeling spatial burstiness,
i.e., high spatial inhomogeneity.

Since the performance characteristics of mobile and, in particular,
wireless systems are strongly influenced by their spatial layout and
topology, point processes and stochastic geometry models have gained
popularity over the last few decades, \seg~\cite{BKLZ_eng:97},
\cite{Zuy:09}, \cite{BacBla:10}, \cite{Hae:12},
\cite{Cou:19}. However, spatial burstiness has received little
attention despite its critical importance for network performance
analysis, dimensioning, and strategic planning. The following example
illustrates this issue.

\begin{figure}[ht] 
\centerline{
\includegraphics[scale=.2]{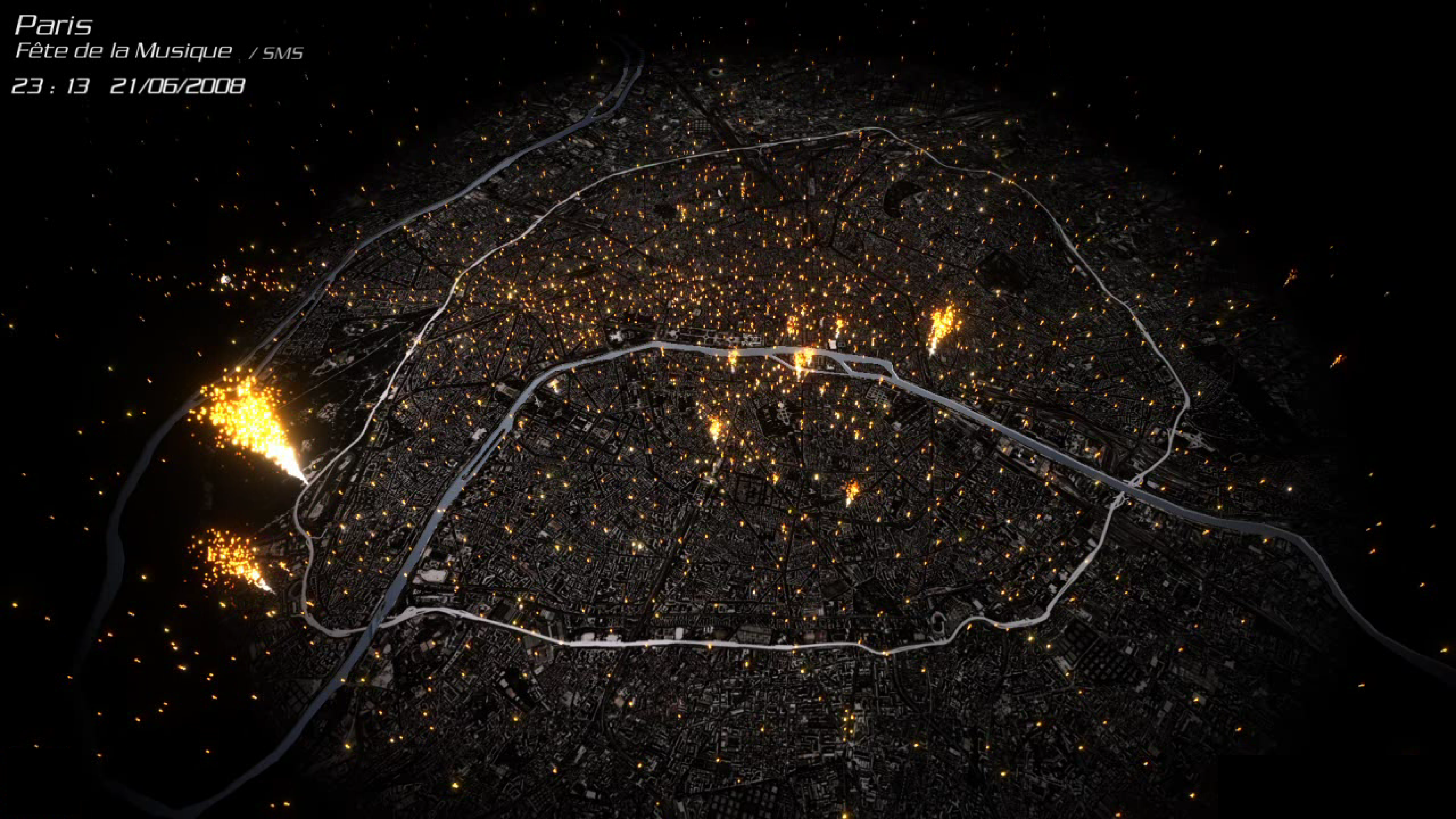}}
\caption{\label{fig:paris} Wireless communication activity during
  \textit{F\^ete de la Musique} festival in Paris in 2008, courtesy of
Orange.fr. The height of the torches represents the
number of calls in progress at their locations.}
\end{figure}

Figure~\ref{fig:paris} presents data on calling activity collected
during the 2008 F\^ete de la Musique festival in Paris.

The height of the torches represents the number of people making calls
in different areas of the city. The data reveal bursty activity,
particularly in locations where large crowds have gathered, such as
stadiums and public squares. When these torches are projected onto the
city’s ground plane, it becomes evident that accurately modeling this
distribution requires a spatial model capable of capturing highly
irregular and dynamic spatial patterns.

This paper develops inference methods for thinning-stable point processes (also known as
discrete-stable processes, or \das\ processes), a generalisation
of the Poisson model that better accommodates bursty spatial patterns.

The structure of the paper is as follows:
Section 2 introduces the concepts of stability and discrete stability
in probability theory and defines thinning-stable (\tas) processes.
Section 3 discusses statistical inference methods.
Section 4 presents simulations and applications.
Section 5 provides concluding remarks.

\section{Mathematical Background}

\subsection{Poisson Limit Theorem}
A Poisson point process is the basic model for point events
happening in time or space. It plays the same role as the Gaussian
processes for continuous data because of the following result mimicking
the Central Limit Theorem. Let $(\Psi_i)$ be
a sequence of \iid~point processes in $\R^d$ with a locally finite intensity
measure $\E \Psi_i(B) =\mu(B)$, \ie~taking finite values for all bounded Borel sets
$B$. Writing $p\circ \Psi$ for the result of \emph{independent thinning} of
the process $\Psi$ points with retention probability $p$, and $+$ for
superposition of the processes, the following \emph{Poisson
limit theorem} holds:
\begin{equation}\label{eq:poislt}
\frac{1}{n} \circ (\Psi_1 + \cdots + \Psi_n) \Rightarrow \Pi_\mu,
\end{equation}
where $\Pi_\mu$ is a Poisson process with intensity measure $\mu$, and
$\Rightarrow$ denotes the weak convergence.

This result explains why Poisson processes naturally emerge as limits
of superpositions of many `thin' sources. The thinning operation is
thus a discrete analogue of the scaling for random variables making
the processes `thin'. However, the assumption of locally finite intensity
measure $\mu$ is inadequate for bursty data, where
variations can span multiple orders of magnitude. For these processes
the expectation $\E\Psi_i(B)$ is infinite even for a bounded $B$ so
the intensity measure does not exist. For such processes, the limit in the 
thinning-superposition scheme is no
longer Poisson. Namely, if for
some $0<\alpha<1$ one has a non-trivial limiting process $\Phi$
such that
\begin{equation}\label{eq:lt}
\frac{1}{n^\alpha} \circ (\Psi_1 + \cdots + \Psi_n) \Rightarrow \Phi,
\end{equation}
then the process $\Phi$ is necessarily \emph{thinning $\alpha$-stable}
or \tas,
\ie\ it satisfies the following distributional identity:
for any $p \in (0,1)$,
\begin{equation}
p^{1/\alpha} \circ \Phi' + (1-p)^{1/\alpha} \circ \Phi'' \deq \Phi,
\end{equation}
where $\Phi', \Phi''$ are independent realisations of $\Phi$ and $\deq$
denotes equality in distribution. Thinning-stable point processes were
introduced and studied in details in \cite{DavMolZuy:11} where we call
them Discrete $\alpha$-Stable (\das). Here we prefer \tas-notation as
it reflects better the used scaling operation $\circ$: a more general
\emph{continuous branching} operation can also lead to a non-trivial limit in
\eqref{eq:lt}, see~\cite{ZanZuy:15a}.

Observe a direct parallel to the classical notion of
stability. Recall that a random variable $\xi$ is strictly
$\alpha$-stable (\sas) if for any $t \in (0,1)$,
\begin{equation}\label{eq:sas}
t^{1/\alpha} \xi' + (1-t)^{1/\alpha} \xi'' \overset{D}{=} \xi,
\end{equation}
where $\xi', \xi''$ are independent copies of $\xi$. Stability plays a
fundamental role in limit theorems, as only stable distributions can
appear as weak limits:
\begin{equation}
n^{-1/\alpha}(\zeta_1 + \cdots + \zeta_n) \Rightarrow \xi.
\end{equation}

Recall that a \emph{Cox process} directed by a random measure $\zeta$
is a double-stochastic Poisson process which may be thought of as been
generated by the following procedure: first, a realisation of
$\zeta(\omega)$ is obtained, second, a realisation of a Poisson
process with $\zeta(\omega)$ as the intensity measure is produced.  By
\cite[Th.~15]{DavMolZuy:11}, \tas\ processes are exactly Cox processes
directed by a random measure $\zeta$ which is \sas, \ie\ for every set
$B$, the random variable $\zeta(B)$ is \sas. Note that Poisson process
is thinning-stable with exponent $\alpha=1$. Thus, for $0<\alpha<1$,
the random measure $\zeta$ plays the same role as a (non-random)
intensity measure of a Poisson process when $\alpha=1$.  An explicit
description of \tas\ process distribution is available in the form of
the probability generating functional, see \cite[Eqns.~(34) and
(40)]{DavMolZuy:11}.

The most practically important class of \tas\ processes comprise the
\emph{regular}\footnote{\tas\ processes are infinitely divisible, so
  \emph{regular} here means the corresponding KLM (or L\'evy) measure
  is supported by the set of finite measures, see
  \cite[Sec.~5]{DavMolZuy:11} for details.} processes which can be
represented as a Poisson cluster processes. This representation allows
direct modeling of burstiness through the underlying Poisson
structure. For this, state a few definitions.

A random variable $\nu$ taking values in $\N$ has Sibuya distribution
$\Sib(\alpha)$ with parameter $\alpha\in(0,1]$ if
\begin{displaymath}
  \P\{\nu=n\} = \prod_{k=1}^{n-1}
  \Big(1-\frac{\alpha}{k}\Big)\frac{\alpha}{n},\ n=1,2,\dots 
\end{displaymath}
It corresponds to the number of Bernoulli trials until the first
success if the probability of success in the $k$-th trial is $\alpha/k$.

A finite point process $\Upsilon$ on a measurable phase space $X$ is called
\emph{Sibuya point process} with parameter $\alpha\in(0,1]$ and a
probability measure $\mu$ on $X$ (notation:
$\Upsilon\sim \Sib(\alpha,\mu)$) if the total number $\nu=\Upsilon(X)$ of
its points is Sibuya $\Sib(\alpha)$-distributed and, given $\nu$,
these points are independently scattered according to the measure
$\mu$.

The main representation theorem for a regular \tas\ process $\Phi$
states that it is characterised by two parameters: $\alpha\in(0,1]$
and a \emph{spectral measure} $\sigma$ which is a measure on the space
$\M_1$ of the probability distributions on $X$. It has the following
\emph{cluster representation}: first, a Poisson process $\widetilde{\Pi}_\sigma$ with
intensity measure $\sigma$ on $\M_1$ is generated (the \emph{germs} process or
cluster \emph{centres}). And then superposition of independent Sibuya
processes (the \emph{daughter} processes or \emph{clusters}) is taken:
\begin{equation}
  \label{eq:clust}
  \Phi \deq \sum_{\mu_i \in \widetilde{\Pi}_\sigma} \Upsilon_i,
\end{equation}
where $\Upsilon_i\sim\Sib(\alpha, \mu_i)$, see~\cite[Th.~24]{DavMolZuy:11}.

\begin{figure}[ht]
  \centerline{
    \includegraphics[scale=.25]{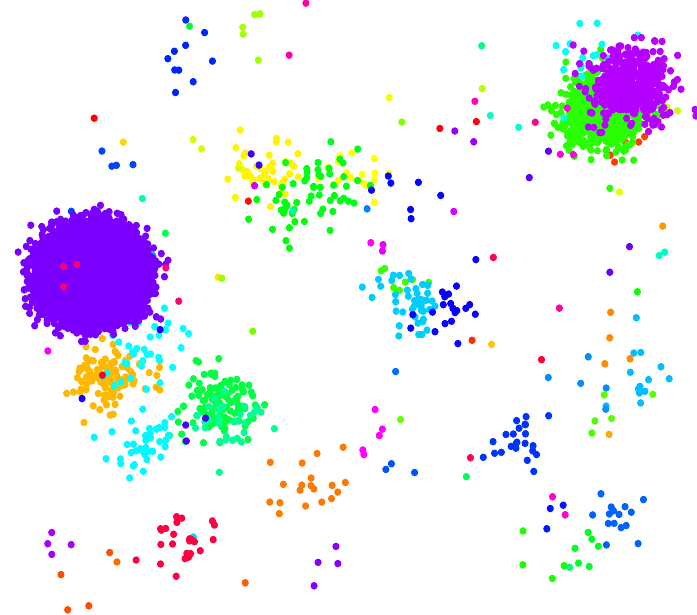}}
% \includegraphics[scale=.35]{../Figures/DaS.png}}
% \caption{$\lambda=0.4$, $\alpha=0.6$, $\sigma=\delta_{\mu_0}$, where
%   $\mu_0 \sim \mathcal{N}(0,0.3^2\mathrm{I})$\label{fig:gauss}}
 \caption{$\lambda=0.4$, $\alpha=0.6$ and
  $\mu_0 \sim \mathcal{MVN}(0,0.5^2\mathrm{I})$\label{fig:gauss}}
\end{figure}

The simplest, yet flexible and rich family of regular \emph{stationary
  ergodic} \tas\ processes on $X=\R^d$ is obtained when the spectral
measure $\sigma$ is concentrated on the set
$\M_{\mu_0}=\{\mu_x = \mu_0(\cdot\,-x):\ x\in\R^d\}$ of shifts of a
fixed measure $\mu_0\in\M_1$ and invariant with respect to these
shifts. The image of such $\sigma$ with respect to mapping
$\mu_x\mapsto x$ from $\M_{\mu_0}$ to $\R^d$ is thus a multiple of the
Lebesgue measure $\lambda dx$ for some $\lambda>0$. Such processes
represent a superposition of independent realisations $\Upsilon_i$ of
$\Sib(\alpha,\mu_0)$-processes shifted to the points of a homogeneous
Poisson process $\Pi_\lambda$ with density $\lambda$ in $\R^d$:
\begin{equation}
  \label{eq:sterg}
  \Phi \deq \sum_{x_i \in \Pi_\lambda} (\Upsilon_i + x_i),
\end{equation}
where
\begin{displaymath}
  \Upsilon_i + x_i = \sum_{y_j\in\Upsilon_i}
\delta_{y_j+x_i}\sim\Sib(\alpha, \mu_0(\cdot\, - x_i)).
\end{displaymath}
This presentation serves a basis for their easy simulations.
Figure~\ref{fig:gauss} shows a realisation of such a process when
$\mu_0$ is a 2D symmetrical Gaussian distribution. Different clusters
are coloured differently for better visibility, but in real data
points may or may not have its cluster identity tag.

\section{Statistical Inference for \tas\ Processes}
For inference, we assume that we observe a stationary ergodic \tas\
process $\Phi$ given by \eqref{eq:sterg} inside a bounded set $W$
(the observation window).

Given an observed realisation of $\Phi$, statistical estimation aims to recover:
\begin{itemize}
\item $\lambda$: the intensity of the cluster centers,
\item $\alpha$: the stability parameter,
\item $\mu_0$: the Sibuya parameter measure, \ie\ the distribution of
  a cluster's points.
\end{itemize}

Since $\Phi$ is a cluster process, we rely on well developed methods
to separate clusters for estimation of $\mu_0$ in Sec.~\ref{sec:A}. In
Sec.~\ref{sec:B}, we propose novel estimation methods for estimation
of $\alpha$ and $\lambda$ which are based on specific stability
property of T$\alpha$S processes with respect to thinning.
    
\subsection{Estimation of the Sibuya parameter measure}\label{sec:A}
The crucial step in statistical inference is estimation of
the measure $\mu_0$ according to which all the clusters are distributed.

One could have three possible situations:
\begin{itemize}
\item $\mu_0$ is already known or assumed; 
\item $\mu_0$ is known to belong to a parametric class (for instance, a
  Gaussian distribution with unknown covariance matrix or a uniform
  distribution in a ball of unknown radius);
\item $\mu_0$ is totally unknown.
\end{itemize}

The bursty nature of \tas\ processes provides a very intuitive way of
estimating $\mu_0$: if one identifies a big cluster and
isolates it from the rest of the data, it would
give an accurate estimation for the measure $\mu_0$ because it contains
a large number of independent points drawn from a shifted version of
$\mu_0$. The center of mass could be taken as an estimate of this shift.

Once a cluster $C$ of a large size $|C|$ is identified, an obvious
estimator of $\mu_0$ is the empirical distribution
\begin{equation}
\widehat{\mu}_0 = \frac{1}{|C|} \sum_{y_i \in C} \delta_{y_i-b},
\end{equation}
where $b = \sum_{y_i \in C} y_i$ is the center of mass. This empirical
measure provides a non-parametric approximation of the spatial
distribution of points in a cluster and serves a basis for subsequent
smoothing, for example, with a kernel density estimation or for
estimation of the parameters via the likelihood maximisation. It is
effective when well-separated clusters can be identified. Then also a
few large clusters could be pooled together after the shift
compensation by their centers of mass $b_k$ providing yet a larger
dataset to estimate $\mu_0$. But detecting cluster may be hard and
unreliable when clusters overlap significantly and the cluster
affiliations are not known.

In these cases, a mixture model provides a better framework for
estimation and it is theoretically supported by the cluster structure
of a \tas\ process. Assuming that $\mu_0(dy)$ has a density, the data $(y_i)$
are interpreted as a realisation from a mixture of
cluster-generating processes with the likelihood
\begin{equation}
\prod_{k} \sum_i p_k f(y_i | \theta_k),
\end{equation}
where $p_k,\ \sum_k p_k=1$, are the probabilities that an
observation point belong to the $k$-th cluster (the mixture weights)
and $f(y \cond \theta_k)$ is the probability density of the $k$-th
Sibuya cluster process parametrised by $\theta_k$.

An Expectation-Maximisation (EM) algorithm can then be applied to
iteratively estimate the parameters:
\begin{itemize}
\item E-step: Compute the expected cluster assignments based on
  current parameter estimates.
\item M-step: Maximize the likelihood function with respect to
  $\theta_k$.
\end{itemize}
This method improves estimation accuracy when the true clusters are
not directly observable, \seg~%\cite{dempster1977maximum} or
\cite{bishop2006pattern}.

\subsection{Estimation of the Stability Parameter and the Cluster Density}
\label{sec:B}
Once an estimate of $\mu_0$ is available, we aim to estimate the other
two parameters: the density $\lambda$ of the clusters' centres and the
stability parameter $\alpha$. Because of availability of explicit
formulas, we tried two methods. One is based on the probability
generating functional (p.g.fl.) which, for a function $u(x)$
such that $0\leq 1-u < 1$ has a bounded support, is given by
\begin{equation}\label{eq:pgfl}
\log G_\Phi[u] = - \int_{\M_1}\langle 1-u, \mu\rangle^{\alpha}
\sigma(d\mu),
\end{equation}
see~\cite[Cor.~16]{DavMolZuy:11}. Here, $\langle f, \mu\rangle$ stands
for $\int_X f(x)\,\mu(dx)$, for short. For the model we consider, this
translates into the log-p.g.f. $\log \E z^{\Phi(B)}$ of the number of points in
a set $B$ which equals:
\begin{equation}
-\lambda (1-z)^\alpha
\int_{\R^d}\mu_0(B-x)^\alpha dx. \label{eq:pgfcount}   
\end{equation}
Fitting empirical p.g.f.\ estimates $\lambda$ and $\alpha$. A boundary
correction is also possible to implement when the integration domain
$\R^d$ is replaced by the observation window $W$.

When there is no multiple points in the data, the process may be
assumed simple (without multiple points). This is the case when
$\mu_0$ has a density. Since void probabilities: $\P\{\Phi(B)=0\}$ for
bounded $B$, determine the distribution of a simple point process
$\Phi$, an alternative method can be used: the parameters are
estimated via fitting the void probabilities for a system of balls of
different radii and positions.  From \eqref{eq:pgfcount}, the void
probability $\P\{\Phi(B) = 0\}$ equals
\begin{equation}\label{eq:void}
\exp\Big\{-\lambda \int_{\R^d} \mu_0(B-x)^{\alpha} dx \Big\},
\end{equation}
% \begin{equation}
% \log\P\{\Phi(B) = 0\}=-\lambda \int_{\R^d} \mu_0(B-x)^{\alpha} dx,
% \end{equation}
and the method is equivalent to fitting the contact distribution tail
function $G(r) = \P\{\Phi(B_r(0))=0\}$, where $B_r(0)$ is the ball of
radius $r$ at the origin. Let $\{x_i\}_{i=1}^n \subset W $ be a
sequence of \textit{test points} 
  and $r_i$ be the distances from $x_i$ to the closest point of
  $\Phi$. It is straightforward to show that
  \begin{equation}\label{eq:g_est}
    \widehat{G}(r)=\frac{1}{n}\sum_{i=1}^{n}\one_{\{r_i>r\}}
  \end{equation}
  is an unbiased estimator for $G(r)$.  Then $\alpha$ and $\lambda$
  are estimated by the best fit to this curve: if $r_{(i)}, 1=1\thru
  n,$ denotes the increasingly ordered sequence of $r_i's$, 
  \begin{align*}
    (\widehat{\alpha}, \widehat{\lambda})&  =
    \underset{\alpha\in(0,1),\, \lambda\geq0}\argmin
                                           \sum_{i=1}^n(\widehat{G}(r_i)-G(r_i))^2 \\
    & = \underset{\alpha\in(0,1),\, \lambda\geq0}\argmin
                                           \sum_{i=1}^n ((n-i)/n-G(r_{(i)}))^2.
  \end{align*}
  It can also be shown that the variance of the estimator is smaller
  when the test points are taken on a grid rather than randomly
  scattered. We also observed poor performance of both estimation
  methods in the presence of large clusters, so for highly bursty
  data. It is explained by the fact that when a test point $x_i$ falls
  inside a large cluster, the distance to the closest point may be by
  orders of magnitude smaller than for the test points in relatively
  void areas. So a novel estimation method is proposed that makes use
  of a particular structure of \tas\ processes.

  The method is based on the observation that a thinned \tas\ process
  is again \tas: it is easy to see from \eqref{eq:pgfl} and that
  $G_{p\circ \Phi}[u] = G_\Phi[1-p(1-u)]$ for independent
  thinning. The parameters $\mu_0$ and $\alpha$ stay the same for
  $p\circ \Phi$, but the clusters' density becomes $\lambda
  p^\alpha$. Therefore,
\begin{align*}
  &\P\{(p \circ \Phi)(B_r(0))=0\}\\
  &=\sum_{k=1}^{\Phi(W)}p(1-p)^{k-1}\P\{r_k>r\}.
\end{align*}
Here $p(1-p)^{k-1}$ is the probability that the closest point of
$\Phi$ which survived the $p$-thinning is the $k$-th. This results in that
    \begin{equation}\label{eq:vp_est}
      \frac{1}{n}\sum_{i=1}^{n}\sum_{k=0} p(1-p)^{k-1}\one_{\{r_{i,k}>r\}}
    \end{equation}
is also an unbiased estimator for $G(r)$. In practice we regress the
values of $\lambda$ and $\alpha$ from the curves obtained for a range
of thinning probability $p$.

\begin{figure}[ht]
  \centering
%   \includegraphics[scale=.32]{realis_0101}
% \end{figure}
% \begin{figure}[ht]
%   \centerline{
%     \includegraphics[scale=.32]{void_prob_est.png}
%   }
   \includegraphics[scale=.21]{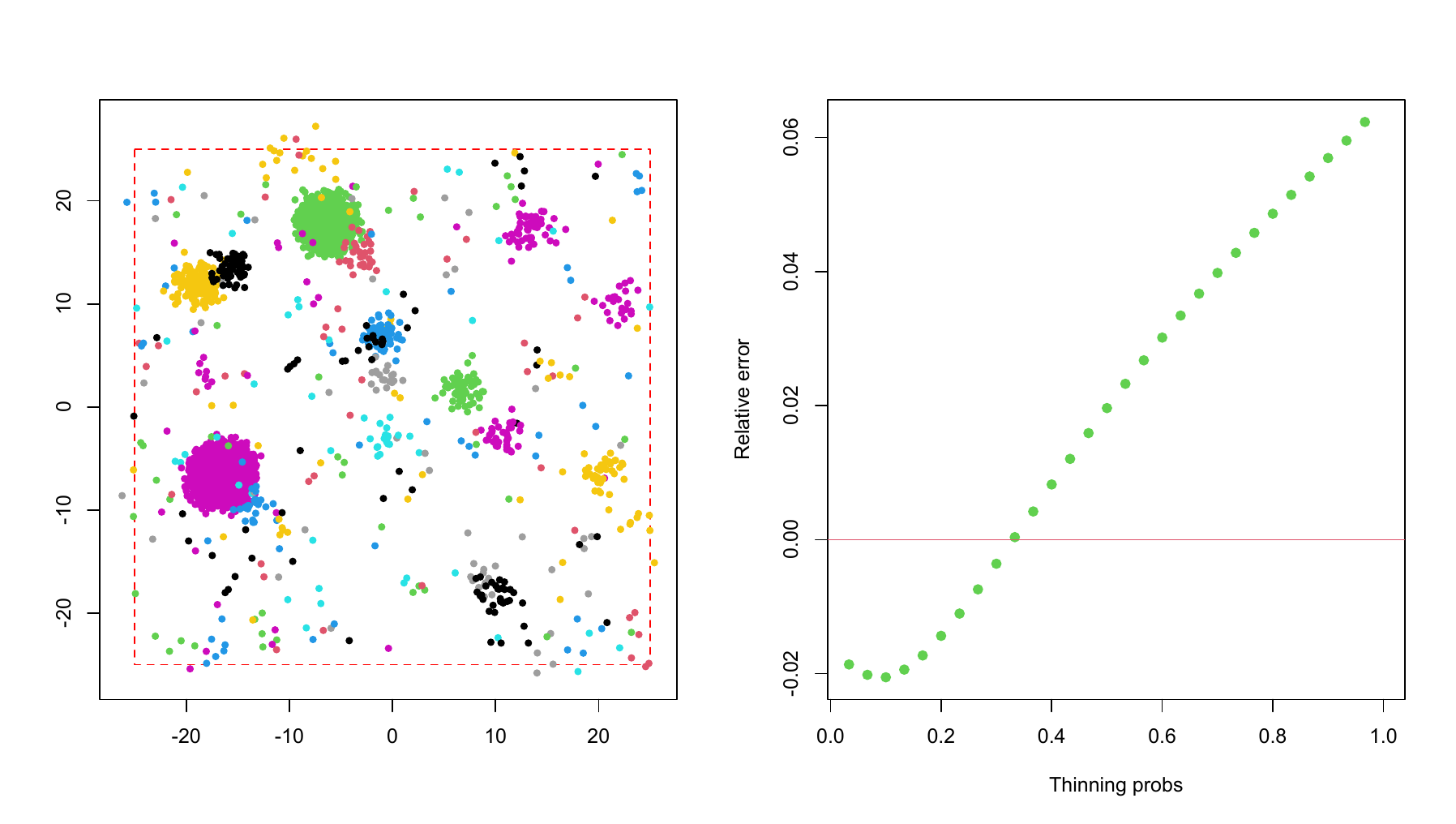}
\caption{Realisation of a Gaussian cluster model with
  $\mu_0\sim\mathrm{MVN}(0,\mathrm{I})$, $\lambda=0.1,
  \alpha=0.7$ and the relative error of
  estimation of $G(1)$ for various values of $p$ in
  \eqref{eq:vp_est}.\label{fig:vp}} 
\end{figure}

The right plot on Figure~\ref{fig:vp} exemplifies the
relative error of the estimation of $G(r)$ for various values of the
thinning probability $p$ for the model with Gaussian clusters
illustrated on the left plot.
Value $p=1$ corresponds to the formula \eqref{eq:g_est}. For small
values of $p$, the error starts to grow because of the gradual loss of
information: a boundary correction starts to prevail because of small
number of points survived such a heavy thinning. We therefore
recommend a moderate range of $p$-values in \eqref{eq:vp_est} which
proves a sufficient number of survived points for the analysis.

\subsection{Estimation of the stability parameter from cluster sizes}
\label{sec:C}
If the clusters are well separated or if they possess an identifying
tag, the stability parameter $\alpha$ may be estimated from their
sizes as these are independent $\Sib(\alpha)$-distributed random
variables. The corresponding p.g.f.\ is given by
$g(t) = 1 - (1-t)^\alpha$, and, given the sizes $n_1\thru n_K$ of $K$
observed clusters we may use an empirical p.g.f.\
$g_K(t) = K^{-1}\sum_{k=1}^K t^{n_k}$ to estimate
\begin{displaymath}
  \alpha = \frac{\log(1-g(t))}{\log(1-t)}.
\end{displaymath}
Fix $0\leq t_1<\dots t_q<1$. Then
\begin{align*}
  \alpha_K & = \sum_{j=1}^q \log(1-g_n(t_j)) b_j,\ \text{where}\\
  b_j & = \frac{l_j - \overline{l}}{\sum_{j=1}^q
        (l_j-\overline{l})^2}\,,\\
  l_j & = \log(1-t_j)\ \text{and}\ 
  \overline{l} = q^{-1} \sum_{j=1}^q l_j
\end{align*}
is a consistent estimator of $\alpha$. Indeed,
\begin{displaymath}
  \alpha_K = \sum_{j=1}^q \frac{\log(1-g_n(t_j))}{l_j} l_j b_j
\end{displaymath}
converges a.s.\ as $K\to\infty$ to
\begin{displaymath}
  \sum_{j=1}^q \frac{\log(1-g(t_j))}{l_j} l_j b_j = \alpha \sum_{j=1}^q
  l_j b_j = \alpha.
\end{displaymath}
Other parameter estimation methods for heavy-tailed distributions
can be employed too.

\section{Applications and Simulations}

We implemented simulation studies to analyse estimation biases under various conditions:
\begin{itemize}
    \item Large vs. moderate datasets;
    \item Overlapping vs. separated clusters;
    \item Heavy burstiness (small $\alpha$, large clusters) vs. moderate burstiness.
\end{itemize}

For the estimation of the distribution $\mu_0$, we observed that
treating some proportion of small clusters as a Poisson noise may
benefit computation time by significantly limiting the possible number
of clusters and hence, possible associations.

To check our estimation methods described in Section~\ref{sec:B}, we
simulated one-dimensional T$\alpha$S process with $\mu_0$ being
Uniform in [-1,1] distribution in the window $W=[-500,500]$ for various
values of parameters ranging from $\alpha=0.6$ (heavy
clusters) to $\alpha=0.8$ (light clusters) and from $\lambda=0.02$
(separated clusters) to $\lambda=0.4$ (overlapping clusters). We
obtained the accuracy
of the estimates are shown in Table~\ref{tab:est} with intermediate
values in between these extremes.
\begin{table}[ht]
  \qquad \qquad \qquad \qquad \qquad \qquad \qquad Cluster density $\lambda$\newline\newline
  \centering
  Stability parameter $\alpha$\quad
  \begin{tabular}[ht]{|c|cc|}
    \hline \hline
    & 0.02 & 0.4\\
    \hline
    0.6 & (0.6045, 0.0208) & (0.6019, 0.3978)\\
    0.8 & (0.8168, 0.0215) & (0.8093, 0.4007)\\
    \hline \hline
  \end{tabular}
  \caption{Estimates $(\widehat{\alpha}, \widehat{\lambda})$
    for the parameters $\alpha$ listed as the row
    labels and $\lambda$ as column labels from 50 realisation of
    T$\alpha$S with uniform $\mu_0$ in 1D.
    \label{tab:est}}
\end{table}
Our simulations show that parameters $\alpha$ and $\lambda$ are best
estimated by void probabilities with thinning method which also
produces best estimates in all the situations apart from moderate
separated clusters, however, at the expense of being more
computationally expensive. The simplest void probabilities method is
well suited for moderate datasets with separated clusters. It
estimates well the burstiness parameter $\alpha$, but in the latter
case $\lambda$ is best estimated by counts' p.g.f.\ fitting. We also
tried fitting $\log G(r)$, because it depends linearly on the
parameter $\lambda$ as seen in \eqref{eq:void}. Then the least squares
approximation error has explicit solution for the root of the partical derivative
with respect to $\lambda$ thus leading to numeric optimisation over just one
parameter $\alpha$. 

As common in modern Statistics, all methods should be tried and
consistency in estimated values gives more trust to the analysis and
the model.

For the F\^ete de la Musique data presented on Figure~\ref{fig:paris},
estimated value of $\alpha$ ranges between 0.17-0.28 for various time
snapshots. Since the clusters are
associated with geographical positions, we employed the method
described in Section~\ref{sec:C}. The obtained estimates indicates a heavy burstiness of
the data and speaks in favour of the use of \tas\ as an appropriate
model. We also observed a poor fit of the Sibuya distribution to small
values of the cluster sizes and tried more general branching-stable
process model \cite{ZanZuy:15a} to improve it.

\section{Conclusion}
Thinning-stable processes provide a powerful framework for bursty
spatial data. Their estimation requires careful selection of inference
methods based on dataset characteristics, specific novel inference methods are
developed in this paper. Further generalisations,
such as branching-stable processes, see \cite{ZanZuy:15a}, can enhance
modeling flexibility.

\section*{Acknowledgements}
\label{sec:acknowledgements}
The author is thankful to his Master students: Brunella Spinelli and Stefano Crespi for
algorithmic implementation in R of the described above methods and for
extensive simulation studies.

% \bibliographystyle{plain}
% \bibliography{papers, mypub, telecom}

\begin{thebibliography}{10}

\bibitem{BKLZ_eng:97}
F.~Baccelli, M.~Klein, M.~Lebourges, and S.~Zuyev.
\newblock Stochastic geometry and architecture of communication networks.
\newblock {\em J. Telecommunication Systems}, 7:209--227, 1997.

\bibitem{BacBla:10}
Fran{\c{c}}ois Baccelli and Bart{\l}omiej B{\l}aszczyszyn.
\newblock {\em Stochastic Geometry and Wireless Networks. Volume 1: Theory}.
\newblock Now Publishers, 2010.

\bibitem{bishop2006pattern}
C.M. Bishop.
\newblock {\em Pattern Recognition and Machine Learning}.
\newblock Springer, 2006.

\bibitem{Cou:19}
David Coupier.
\newblock {\em Stochastic Geometry Modern Research Frontiers: Modern Research
  Frontiers}.
\newblock Springer, 01 2019.

\bibitem{DavMolZuy:11}
Yu. Davydov, I.~Molchanov, and S.~Zuyev.
\newblock Stability for random measures, point processes and discrete
  semigroups.
\newblock {\em Bernoulli}, 17(3):1015--1043, 2011.

\bibitem{erlang1909}
A.K. Erlang.
\newblock {S}andsynlighedsregning og telefontrafik.
\newblock {\em Tidsskrift for Matematik B}, 20, 1909.

\bibitem{Hae:12}
Martin Haenggi.
\newblock {\em Stochastic Geometry for Wireless Networks}.
\newblock Cambridge University Press, 2012.

\bibitem{Kl_roc:75}
L.~Kleinrock.
\newblock {\em Queueing systems. Volume I: Theory}.
\newblock Wiley, 1975.

\bibitem{palm1943}
C.~Palm.
\newblock {I}ntensit\"{a}tsschwankungen im fernsprechverkehr.
\newblock {\em Ericsson Technics}, 1943.

\bibitem{ZanZuy:15a}
G.~Zanella and S.~Zuyev.
\newblock Branching-stable point processes.
\newblock {\em Electronic J. Probab.}, 20(119):1--26, 2015.

\bibitem{Zuy:09}
S.~Zuyev.
\newblock Stochastic geometry and telecommunications networks.
\newblock In Kendal.~W. S. and I.~Molchanov, editors, {\em Stochastic Geometry:
  Highlights, Interactions and New Perspectives}, pages 520--554. Oxford
  University Press, 2009.

\end{thebibliography}

\end{document}